\documentclass{ws-procs9x6}

\makeatletter
\def\simleq{\mathrel{\mathpalette\gl@align<}}
\def\simgeq{\mathrel{\mathpalette\gl@align>}}
\def\gl@align#1#2{\lower.6ex\vbox{\baselineskip\z@skip\lineskip\z@
     \ialign{$\m@th#1\hfill##\hfil$\crcr#2\crcr\sim\crcr}}}
\makeatother

\newcommand{\bra}{\langle}
\newcommand{\ket}{\rangle}
\newcommand{\braket}[1]{\bra #1 \ket}
\newcommand{\qq}{\braket{\bar{q}q}}

\newcommand{\qGq}{g\braket{\bar{q}\sigma_{\mu\nu}G_{\mu\nu} q}}

\begin{document}

\title{Lattice QCD study for\\
the Quark-Gluon Mixed Condensate 
$\lowercase{g}\langle\bar{\lowercase{q}}\sigma_{\mu\nu}G_{\mu\nu} \lowercase{q}\rangle$}

\author{Takumi Doi$^1$, Noriyoshi Ishii$^2$, Makoto Oka$^1$ and \\
	Hideo Suganuma$^1$}

\address{$^1$ Dept. of Phys., 
         Tokyo Institute of Technology, 
         Meguro, Tokyo 152-8551, Japan\\
	 E-mail: doi@th.phys.titech.ac.jp}

\address{$^2$ Radiation Laboratory, RIKEN, 
	Hirosawa 2-1, Wako, 351-0198, Japan}


\maketitle

\abstracts{
We study the quark-gluon mixed condensate 
$g\langle\bar{q}\sigma_{\mu\nu}G_{\mu\nu} q\rangle$, which is another chiral order parameter,
using the SU(3)$_c$ lattice QCD with the Kogut-Susskind(KS) fermion at the quenched level.
We generate 100 gauge configurations on the $16^4$ lattice with $\beta = 6.0$, and 
measure $g\langle\bar{q}\sigma_{\mu\nu}G_{\mu\nu} q\rangle$ 
at 16 points in each gauge configuration for each current-quark mass of $m_q=2
1, 36, 52$ MeV.
From the 1600 data for each $m_q$, 
we find $m_0^2 \equiv 
g\langle\bar{q}\sigma_{\mu\nu}G_{\mu\nu} q\rangle / 
\langle\bar{q}q\rangle
\simeq 2.5$ GeV$^2$ 
at the lattice scale of $a^{-1} \simeq 2 {\rm GeV}$ in the chiral limit.
The large value of 
$g\langle\bar{q}\sigma_{\mu\nu}G_{\mu\nu} q\rangle$
suggests its importance in the operator product expansion(OPE) in QCD.}
%

\vspace*{-8mm}

\section{The importance of the quark-gluon mixed condensate}
\label{sec:intro}

In order to understand the non-perturbative structure
of the QCD vacuum, condensates are important quantities.
%
%
Among various condensates, we emphasize the importance 
of the quark-gluon mixed condensate 
$\qGq
$.
First, 
the mixed condensate represents a direct correlation 
between quarks and gluons in the QCD vacuum even at the qualitative level.
%
%
Second, the mixed condensate is another chiral order parameter,
which flips the chirality of the quark.
Third, the mixed condensate plays an important role in
various QCD sum rules, especially in the baryons\cite{Dosch},
the light-heavy mesons\cite{Dosch2}
and the exotic mesons\cite{Latorre}.
In the QCD sum rules,
the value $m_0^2 \equiv \qGq / \qq \simeq 0.8\pm 0.2\ {\rm GeV}^2$ has been 
proposed 
phenomenologically\cite{Bel}.
However, 
there was only one pioneering but preliminary
lattice QCD work\cite{K&S},
performed with very little 
statistics (only 5 data)
using a small ($8^4$) 
and coarse lattice ($\beta=5.7$).
Therefore, we present the calculation for $\qGq$ in 
the SU(3)$_c$ lattice QCD with a larger $(16^4)$ and finer $(\beta=6.0)$
lattice and with high statistics (1600 data),
using the KS-fermion at the quenched level.
With this high statistics, 
we perform reliable estimate for 
 $m_0^2\equiv \qGq/\qq$ at the lattice scale 
in the chiral limit.

\vspace*{-3mm}
\section{Formalism and the Results of Lattice QCD}
\label{sec:formalism}

We calculate the condensates $\qq$ and $\qGq$ 
using the SU(3)$_c$ lattice QCD with the KS-fermion 
at the quenched level.
The KS-fermion preserves the explicit chiral symmetry
for the quark mass $m=0$. This is essential for our study
because both of $\qq$  and  $\qGq$
work as the chiral order parameter.
We generate 100 gauge configurations with the standard 
Wilson action at $\beta=6.0$ on the $16^4$ lattice.
The lattice unit\cite{Takahashi} $a\simeq 0.10{\rm fm}$ is obtained
so as to reproduce 
the string tension $\sigma = 0.89 {\rm GeV/fm}$.
We use $m = 21, 36, 52$ MeV
(i.e. $ma = 0.0105,\ 0.0184,\ 0.0263$).
%
%
%
%
%
%
We calculate the flavor-averaged condensates as 
%
\begin{eqnarray}
&& a^3 \qq 
= - \frac{1}{4}\sum_f {\rm Tr}\left[ \braket{q^f(x) \bar{q}^f(x)} \right], \\
&&a^5 \qGq
= - \frac{1}{4}\sum_f \sum_{\mu,\nu}{\rm Tr}
	\left[ \braket{q^f(x) \bar{q}^f(x)} \sigma_{\mu\nu} G_{\mu\nu}\right],
%
\end{eqnarray}
where SU(4)$_f$ quark-spinor field $q$, $\bar{q}$ is converted into 
spinless Grassmann KS-fields $\chi$, $\bar{\chi}$
and the gauge-link variable\cite{DOIS:qGq}.
%
%
%
%
%
We adopt the clover-type definition of
the gluon field strength $G_{\mu\nu}$ on the lattice as 
\begin{eqnarray}
\label{eq:clover}
G_{\mu\nu}^{\rm lat}(s) &=& \frac{i}{16} \sum_A 
\sum_{\stackrel
	{\mbox{$\scriptstyle s'= s,s-\mu,$}}
	{\mbox{$\scriptstyle \ \ \ \ \ s-\nu,s-\mu-\nu$}}}
\lambda^A \ {\rm Tr}
\left[\ 
 \lambda^A\{ U_{\mu\nu}(s') - U_{\nu\mu}(s') \} 
\ \right],
%
%
\end{eqnarray}
which has no ${O}(a)$ discretization error. 
%
%
%
We measure the condensates 
on 16 different space-time points $x$ in each configuration.
%
%
For each $m$, we take the average over the 16 space-time points and 
100 gauge configurations.

\begin{figure}[hbt]
\includegraphics[scale=0.4]{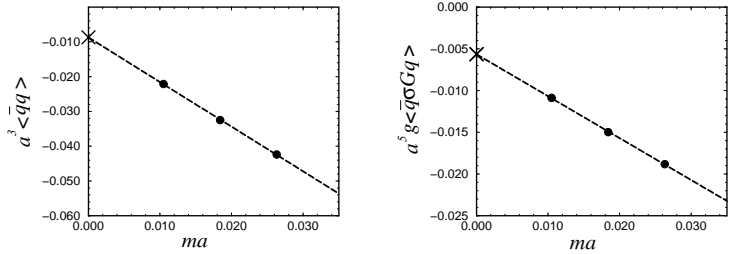}
\vspace*{-3mm}
\caption{\label{fig:plot}
The bare condensates $\qq$ and $\qGq$ plotted against the quark mass $ma$.
The dashed lines denote the best linear extrapolations,
and the cross symbols the values in the chiral limit. 
The jackknife errors are hidden in the circles.
}
\vspace*{-2mm}
\end{figure}

Figure~\ref{fig:plot} shows the 
bare condensates $\qq$ and $\qGq$ against the quark mass $ma$.
Note that the jackknife errors are almost negligible
due to the high statistics of $1600$ data for each $m$.
We fit the data  linearly and determine 
the condensates in the chiral limit.
The obtained data are summarized in Table~\ref{tab:mass-beta-6.0}.
We also check the finite volume artifact,
by imposing the anti-periodic/periodic boundary conditions on $\chi, \bar{\chi}$.
The two results with different boundary conditions almost coincide within 
about 1\% deviation and 
we conclude that the physical volume
$V \sim (1.6\ {\rm fm})^4$ in our simulations is large enough to 
avoid the finite volume artifact\cite{DOIS:qGq}.
%


\begin{table}[hbt]
\tbl{The numerical results of $\qq$ and $\qGq$ for various $ma$.
The last column denotes their values in the chiral limit by the 
chiral extrapolation.\vspace*{1pt}}
{\footnotesize
\begin{tabular}{|c|c|c|c|c|}
\hline
	   &  $ma=0.0263$     &  $ma=0.0184$     &  $ma=0.0105$      &  chiral limit\\
\hline
$a^3\qq$   &  $-0.042397(16)$ &  $-0.032470(15)$ &  $-0.022124(16)$  & 
$-0.008721(17)$ \\
$a^5\qGq$  &  $-0.018820(15)$ &  $-0.014979(14)$ &  $-0.010884(14)$  & 
$-0.005652(14)$ \\
\hline
\end{tabular}}
\label{tab:mass-beta-6.0}
\vspace*{-5pt}
\end{table}


The values  of the  condensates in the  continuum limit  
are to be obtained through the renormalization,
which, however, suffers from
uncertainty of the non-perturbative effect.
As a more reliable quantity,
we provide   the  ratio  $m_0^2  \equiv  \qGq   /  \qq$,
which is free  from the  uncertainty from  the wave
function  renormalization  of the  quark.
Now, we present the result of $m_0^2$ using our bare results
of lattice QCD as
%
%
%
%
\begin{eqnarray}
m_0^2 \equiv \qGq / \qq  \simeq  2.5\ {\rm GeV}^2 \qquad (\beta = 6.0 \ {\rm or} \ a^{-1} \simeq 2{\rm GeV}).
\end{eqnarray}
We see that $m_0^2$ is  rather large, which suggests the importance of
the   mixed  condensate   in  OPE.
%
%
Note that this  bare result
is determined very precisely\cite{DOIS:qGq}.


Finally, 
we  again  emphasize  that  the  mixed condensate  $\qGq$  plays  very
important roles  in various contexts in quark  hadron physics. 
In particular,
the thermal effect is interesting 
because the  mixed  condensate is  another chiral  order
parameter. 
Considering the importance of finite-temperature QCD in the RHIC project, 
we are in progress
for the study of the mixed condensate $\qGq$ at finite temperature\cite{DOIS:T}.



\vspace*{-3mm}

\begin{thebibliography}{0}



\bibitem{Dosch}     {H.G. Dosch, M. Jamin and S. Narison,
                        {\it Phys. Lett.} {\bf B220}, 251 (1989).}

\bibitem{Dosch2}     {H.G. Dosch and S. Narison,
                        {\it Phys. Lett.} {\bf B417}, 173 (1998)}

\bibitem{Latorre}    {J.I. Latorre, P. Pascual and S. Narison,
                        {\it Z. Phys.} {\bf C34}, 347 (1987).}

\bibitem{Bel}       {V.M. Belyaev and B.L. Ioffe,
                        {\it Sov. Phys. JETP} {\bf 56}, 493 (1982).}




\bibitem{K&S}       {M. Kremer and G. Schierholz,
                        {\it Phys. Lett.} {\bf B194}, 283 (1987).}

\bibitem{Takahashi}    {T.T. Takahashi et al.,
			{\it Phys. Rev.} {\bf D65}, 114509 (2002).}

\bibitem{DOIS:qGq}  {T. Doi, N. Ishii, M. Oka and H. Suganuma,
			hep-lat/0211039 (2002).}

\bibitem{DOIS:T}    {T. Doi, N. Ishii, M. Oka and H. Suganuma,
			hep-lat/0212006 (2002).}





\end{thebibliography}
\end{document}